\documentclass[preprint,english,aps,showpacs]{revtex4-1}

\usepackage{graphicx}

\usepackage{babel}

\begin{document}

\title{Macro- and micro-strain in GaN nanowires on Si(111)}
\author{B. Jenichen}
\email{bernd.jenichen@pdi-berlin.de}
\author{O.~Brandt}
\author{C.~Pf\"uller}
\author{P.~Dogan}
\author{M.~Knelangen}
\author{A.~Trampert}
\affiliation{Paul-Drude-Institut fuer Festkoerperelektronik,
Hausvogteiplatz 5--7,D-10117 Berlin, Germany}

\date{\today}

\begin{abstract}
We analyze the strain state of GaN nanowire ensembles by x-ray diffraction.
The nanowires are grown by molecular beam epitaxy on a Si(111) substrate in
a self-organized manner. On a macroscopic scale, the nanowires are found
to be free of strain. However, coalescence of the nanowires results in
micro-strain with a magnitude from $\pm(0.015)\%$ to $\pm(0.03)\%$.
This micro-strain contributes to the linewidth observed in low-temperature
photoluminescence spectra.
\end{abstract}


\maketitle

\section{Introduction}
Epitaxially grown semiconductor nanowires (NWs) have attracted large
interest in recent years as means to fabricate nano-sized devices. They
have been grown using various techniques like e.g.~molecular beam
epitaxy (MBE) and metal organic chemical vapor phase epitaxy.
Often metal catalyst particles are used to initiate the growth.
More recently, catalyst-free processes have been developed.
The NWs promise to assist in overcoming many of the
limitations of heteroepitaxy.  Dislocations formed at the NW/substrate
interface do not propagate along the NW axis, but remain at the interface
or bend toward the free NW sidewalls.\cite{harui_jjap_08,consonni_prb_10}
Residual strain is released elastically due to the extreme aspect ratio
of typical semiconductor NWs.\cite{ertekin_jap_05,glas_prb_06} As a
specific example, GaN NWs on Si are virtually free of extended defects
despite the very large lattice and thermal mismatch between these
materials.\cite{calleja_prb_00} Furthermore, the spectral position of
the donor-bound exciton [(D$^0$,X$_A$)] transition in photoluminescence
(PL) spectra of GaN NW ensembles is exactly the same as the one observed
in  strain free, bulk GaN.\cite{brandt2010} In apparent contradiction,
the linewidth of this transition exceeds values of 1~meV and is thus
significantly above the value expected for material free of
strain.\cite{nostrand_jcg_06,furtmayr_jap_08,brandt2010} Recently,
this minimum broadening invariably observed in the PL of GaN NW ensembles
has been proposed to arise from the random position of donor sites and
the corresponding energy distribution of excitons bound to these
donors.\cite{corfdir_jap_09,brandt2010} However, this intrinsic
broadening mechanism cannot account for the fact that the linewidth
observed for different samples may vary significantly, and may also
approach values in excess of the energy separation of the donor-bound
exciton transition in the bulk and close to the surface.

Here, we investigate the morphological and structural properties of
two GaN NW ensembles on Si(111) exhibiting a PL linewidth differing
by more than a factor of two. Using x-ray diffraction (XRD), we determine the
orientation distribution of the NW ensembles as well as their strain
state. Position and  shape of the x-ray diffraction peaks
contain information about the strains and sizes as well as the
orientation distribution of the NW ensemble. Both samples are found
to be entirely relaxed on a macroscopic
scale, but to possess a non-negligible micro-strain.

\section{Experiment}

GaN NWs are grown on Si(111) substrates by plasma-assisted MBE.
Atomic N and Ga atoms are supplied by a plasma source
operating at 500~W and an N$_2$ flux of 2~sccm. Prior to NW growth,
the Si substrate is exposed for 5~min to an atomic N flux for
nitridation, which results in the formation of a 5~nm thick Si$_x$N$_y$
film. For sample~1 (2), GaN NWs were subsequently grown for 3~h at a
substrate temperature of $780~(820)^\circ$C and N rich conditions,
resulting in NW arrays with an average length of 1.0 (1.6)~$\mu$m.\cite{Dogan2011}
High-resolution XRD measurements were performed at a temperature of
$27^\circ$C using a Panalytical X-Pert PRO MRD\texttrademark\ system
with a Ge(220) hybrid monochromator  (CuK$\alpha_1$ radiation with a
wavelength of $\lambda=1.54056$~\AA). The circles of the diffractometer
are explained in Ref. \onlinecite{Fewster95}. The program
Epitaxy\texttrademark\ was used for evaluation of the data. PL spectroscopy
was performed using the 325-nm line of a Kimmon He-Cd laser for excitation.
The PL signal was dispersed by an 80-cm Horiba Jobin-Yvon monochromator.
The spectral resolution of  the setup was about 250~$\mu$eV.
The investigations were done at a temperature of 10~K.
TEM specimens were prepared by mechanical lapping and polishing, followed by
Argon ion milling. TEM images were acquired with a JEOL 3010 microscope
operating at 300 kV.

\section{Results and Discussion}

\begin{figure}[!t]
\includegraphics[width=6.0cm]{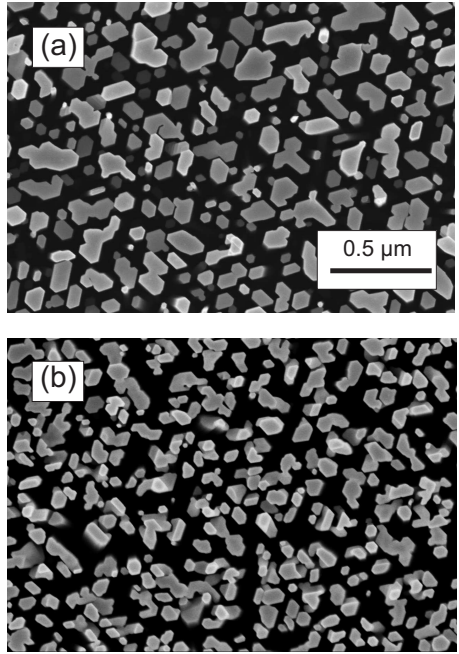}
\caption{Scanning electron micrographs of (a) sample~1 and (b) sample~2
in top view. High-density arrays of facetted NWs are visible. The regular
orientations of the facets point to a narrow distribution of the azimuthal
orientation of the NWs. A significant fraction of the NWs constituting the
ensembles have undergone coalescence.}
\label{fig:sem1}
\end{figure}

Figures~\ref{fig:sem1}(a) and (b) display top-view scanning electron
micrographs of samples~1 and~2, respectively. The azimuthal orientation
of the NWs, as seen by the direction of their facets, is quite regular
although the interfacial Si$_x$N$_y$ layer is not expected to grow
epitaxially. The NWs of sample~2 are distinctly thinner compared to
those of sample~1. Coalescence of the NWs is obvious for both samples.

\begin{figure}[!b]
\includegraphics[width=6cm]{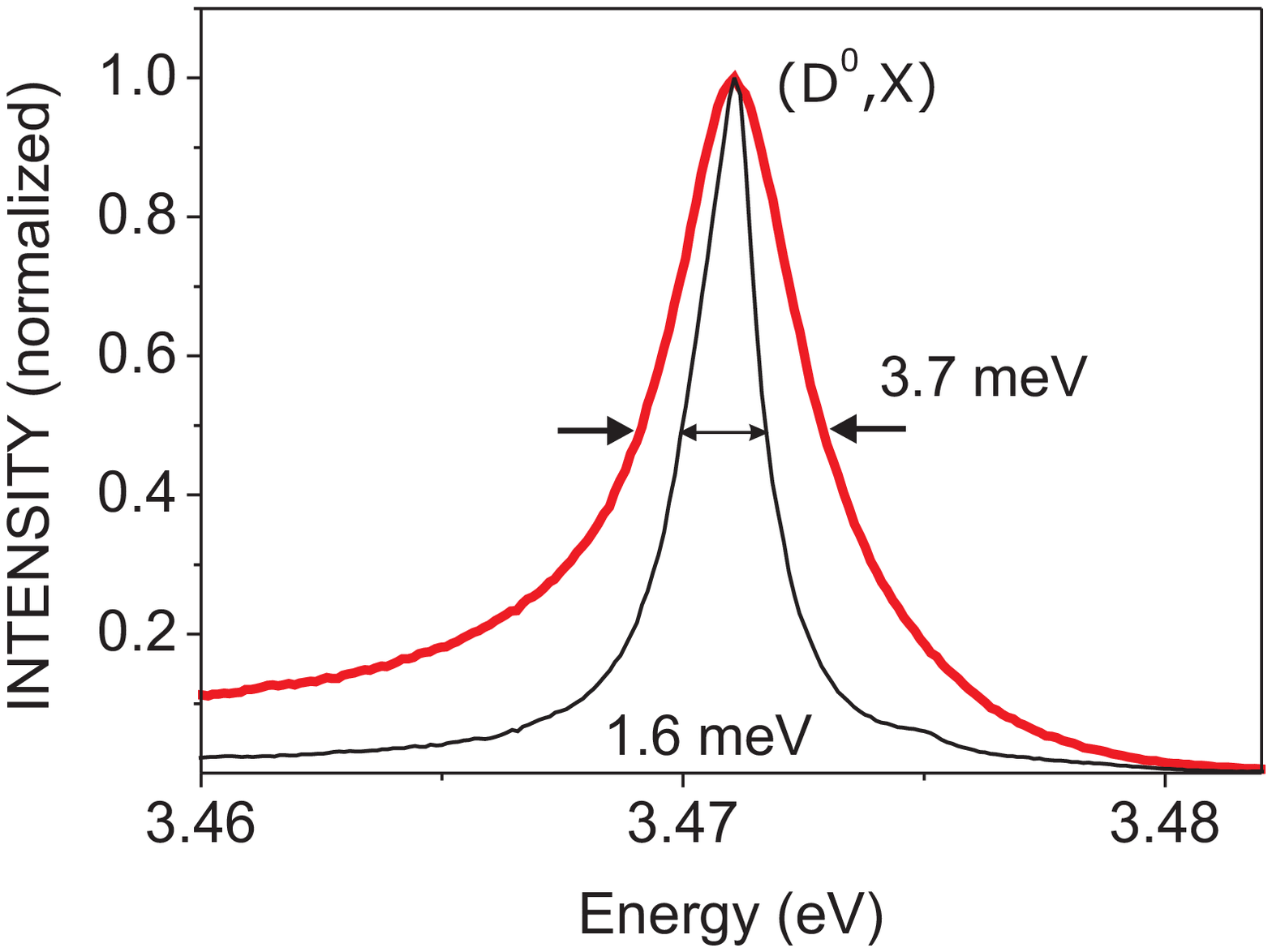}
\caption{(Color online) Near-bandgap PL spectra of samples~1 (thick line)
and~2 (thin line) at 10~K. The dominant line at 3.471~eV has a linewidth
of 3.7~meV~(1.6~meV) for sample~1~(2) and is attributed to the decay of
the donor-bound exciton in strain-free GaN.}
\label{fig:pl}
\end{figure}

Figure~\ref{fig:pl} shows near-bandgap PL spectra of the NW ensembles displayed
in Fig.~\ref{fig:sem1}. For both samples, the dominant line at 3.471~eV is due
to the decay of the donor-bound exciton in strain-free
GaN.\cite{kornitzer_pssb_99,wysmolek_prb_02}
Sample~1 exhibits a linewidth of 3.7~meV, more than twice larger than the one
observed for sample~2 (1.6~meV). Broadening of bound exciton transitions by
their energy dispersion as a function of the distance to the surface is expected
to be more important for thin NWs, since the fraction of donors close to the
surface increases quadratically with the diameter.\cite{corfdir_jap_09,brandt2010}
In the present case, we would thus expect a narrower transition for sample~1
and not a broader one as observed experimentally. Consequently, the additional
broadening observed for this sample has a different origin. As hypothesized in
the following, a highly probable source of this broadening is microstrain
caused by coalescence.

The two samples under investigation were fabricated at different substrate
temperatures, which is known to affect both the nucleation density and the
lateral growth rate. Lower temperatures as used for sample~1 result in a higher
nucleation density as well as a more pronounced tendency towards lateral growth
during nucleation, thus resulting in larger nuclei. This scenario will inevitably
lead to an enhanced coalescence in an early stage of NW growth, and significantly
larger diameters of the final NWs as indeed observed in Fig.~\ref{fig:sem1}.
The misorientation of the NWs with respect to each other, as well as the dislocations
accommodating this misorientation at the tilt or twist boundary,\cite{Consonni2009}
will introduce an inhomogeneous elastic distortion of the coalesced aggregate.
Simultaneously, the strain introduced at the coalescence junction is more difficult
to relax in the thicker  NWs of sample~1.\cite{glas_prb_06} All these arguments
would point towards a larger contribution of inhomogeneous strain to the PL
linewidth of sample~1, as also observed experimentally. In the following, we
examine this hypothesis using high-resolution XRD.

\begin{figure}[tbp]
\includegraphics[width=8cm]{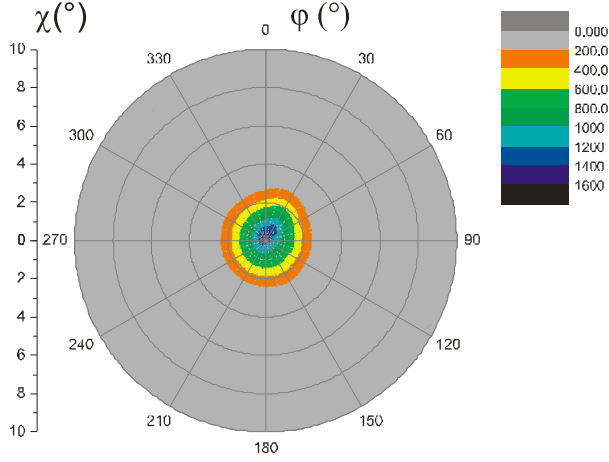}
\caption{(Color online) X-ray pole figure of the GaN 00.4 reflection (sample~1) showing a nearly isotropic
angular distribution of the tilt of the NWs. In this measurement $\chi$ is restricted
to $0^\circ\leq\chi\leq10^\circ$.}
\label{fig:pole2}
\end{figure}

\begin{figure}[tbp]
\includegraphics[width=8cm]{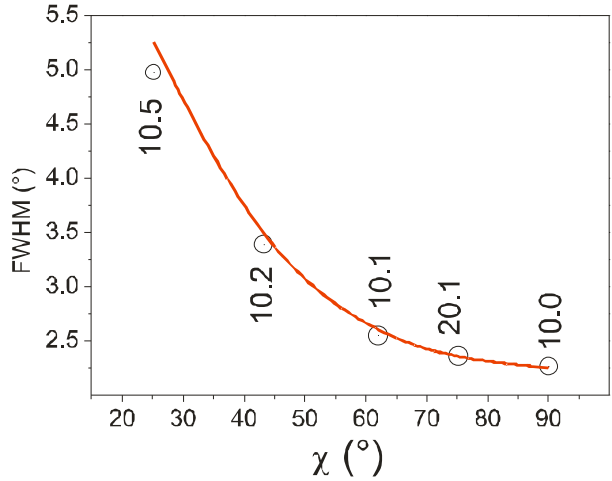}
\caption{(Color online) Dependence of the full widths at half maximum (circles) of of the maxima of  x-ray $\varphi$-scans
using different reflections (sample~1), each tilted by a certain angle $\chi$ with respect to the sample surface.
FWHMs extrapolated by Eq.~(\ref{eq1}) are shown as well (full line).}
\label{fig:twist}
\end{figure}

The out-of-plane orientation distribution of the NWs is determined from
$\omega$ scans near the GaN 00.4 reflection at several azimuthal orientations.
The tilt of $(2.3\pm0.2)^\circ$ [$(3.2\pm0.1)^\circ$] for sample~1 [2] are
typical for GaN NWs grown directly on Si(111). A pole figure of the GaN 00.4 reflection
(sample~1) shown in Fig.~\ref{fig:pole2} displays a nearly isotropic angular distribution of the tilt angles
of the GaN NWs analogous to a fibre texture.
Figure~\ref{fig:twist} demonstrates the dependence of the full widths at half maximum (FWHM) of the
peaks of $\varphi$-scans (circles) performed using different reflections in skew geometry (sample~1)
at different tilt angles $\chi$ (see \cite{Sun2002} for a sketch of the skew geometry)  including
in-plane scans for determination of the range of twist of the NWs. The full line shows the FWHMs extrapolated by:

\begin{equation}
W_{h,k,l}(\chi) =  W_{90}/(W_{90}/360 + \sin(\chi))
\label{eq1}
\end{equation}

 where $W_{h,k,l}(\chi)$ is the FWHM of a peak in the  $\varphi$-scan for a reflection ${h,k,l}$ with a given
 tilt angle $\chi$ , and $W_{90}$ is the FWHM of a peak in an in-plane scan revealing the true range of twist
 of the NWs. Often a $\varphi$-scan can be measured in skew geometry more easily than in an in-plane
 geometry, then $W_{h,k,l}(\chi)$ is measured and $W_{90}$ can be obtained from Eq.~(\ref{eq1}).

\begin{figure}[tbp]
\includegraphics[width=7cm]{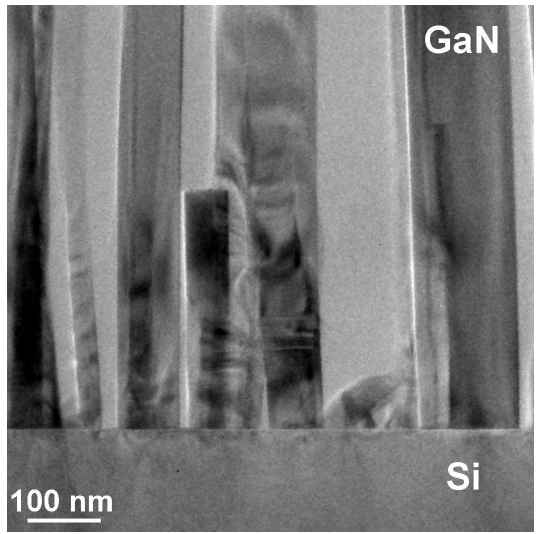}
\caption{TEM micrograph of some of the GaN NWs on top of the
Si$_{x}$N$_{y}$-film on Si 111 (sample~1). The local distribution of strain
is visible from the contrasts on the micrograph. Those strain contrasts
mainly due to bending of the NWs are found near the bottom of the NWs and
near the coalesced regions. }
\label{fig:tem}
\end{figure}

\begin{figure}[tbp]
\includegraphics[width=7cm]{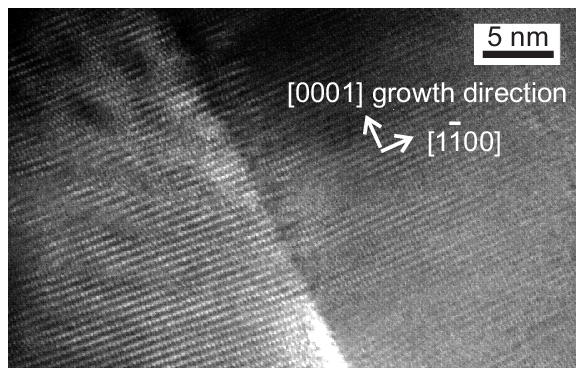}
\caption{High-resolution TEM micrograph of the coalesced region of two GaN NWs (sample 1).
Dislocations are formed at the border similar to a small angle grain boundary and the netplanes are curved locally due to
the strain field of the defects. }
\label{fig:tem2}
\end{figure}

\begin{figure}[tbp]
\includegraphics[width=7cm]{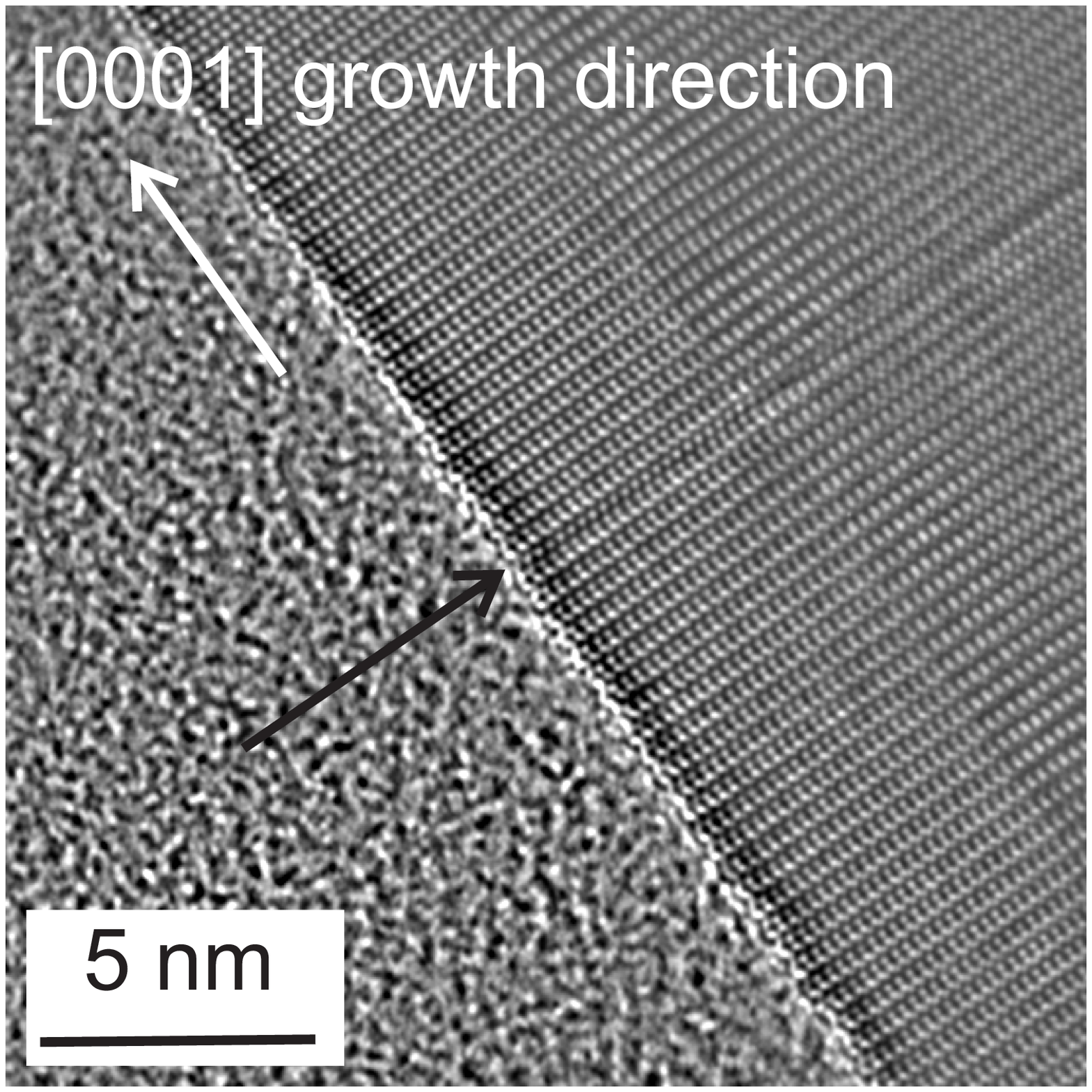}
\caption{High-resolution TEM micrograph (Fourier filtered) of an isolated crystal defect (see black arrow) in a GaN NW (sample 1).
This defect is probably caused by a stacking fault bounded by a dislocation. }
\label{fig:tem3}
\end{figure}

The in-plane orientation
distribution of the NWs as obtained from $\varphi$ scans of the GaN 10.0
reflection is about $(2.3\pm0.3)^\circ$ [$(3.9\pm0.3)^\circ$] for
sample~1 [2], and is again characteristic for GaN NWs on Si(111).
For both samples, coalescence boundaries will be the source of significant
strain due to the broad orientation distribution of the NWs.  This is
illustrated in Figures~\ref{fig:tem} and ~\ref{fig:tem2}: Fig.~\ref{fig:tem} shows a TEM micrograph of the interface
region of sample~1, the GaN NWs on the Si$_{x}$N$_{y}$ intermediate film.
Strain contrasts are clearly distinguished near the bottom of the NWs  and
in regions where NWs meet and coalesce. Fig.~\ref{fig:tem2} demonstrates a
high-resolution TEM micrograph of the coalesced region of two GaN NWs (sample 1).
The border acts as a small angle grain boundary and dislocations are formed in the vicinity
to accommodate the tilt. The netplanes are locally curved due to the strain field of the defects.
Fig.~\ref{fig:tem3} shows as another example a high-resolution TEM micrograph of an isolated crystal defect in a GaN NW (sample 1).
This defect is probably caused by a stacking fault bounded by a dislocation.

To determine the macroscopic strain state of the NWs, we first measured the
lattice parameter of our Si substrates by triple-crystal
diffractometry \cite{Fewster95,moram_rpp_09}. Using an analyzer crystal
in front of the detector, the angular positions of the incident and the
diffracted beams are determined precisely in order to obtain the
Bragg angle. The necessary calculations for the substrate are  performed
in dynamical approximation \cite{Stepanov1997}. We thus obtained a
value of  $5.4299$~\AA. Thanks to the small diameter of the NWs, the
kinematical approximation is sufficient for the calculation of the angular
position of their x-ray reflections. The positions of the symmetrical
in-plane and out-of-plane reflections in $\omega/2\Theta$ scans yield
$a=3.1887$ (3.1885)~\AA\ and $c = 5.1855$ (5.1854)~\AA\ for the lattice
parameters of sample~1 (2). Within the error margin of
$\pm~5\times10^{-4}$~\AA, these values agree with those reported in
Refs.~\onlinecite{moram_rpp_09} and  \onlinecite{Robins2007} for
strain-free and pure GaN. This means, that our GaN NWs are indeed free of
strain on average.

\begin{figure}[!t]
\includegraphics[width=6cm]{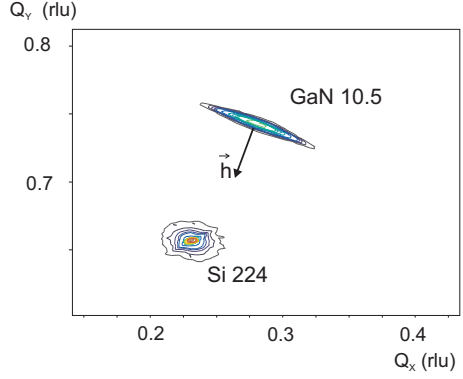}
\caption{(Color online) Reciprocal space map near the GaN 10.5 and the
Si 224 reflection for sample~1. The GaN lattice is fully relaxed, i.e.,
the reciprocal lattice vector of the diffracting planes $h$ (arrow) is
pointing to the origin of reciprocal space. rlu$=\lambda/(2d)$ stands for
reciprocal lattice unit.}
\label{fig:RSM}
\end{figure}
The reciprocal space map shown in Fig.~\ref{fig:RSM} for sample~1 confirms
the fully relaxed state of the GaN NWs, because the diffraction vector
$\overrightarrow{h}$ is pointing in radial direction from the GaN 10.5
peak towards the origin of reciprocal space. The mosaic spread of the
GaN NWs can be calculated from the width of the GaN peak perpendicular
to the radial direction. The lateral correlation length is calculated
from the reciprocal of the FWHM of the GaN~10.5 peak measured parallel
to the interface, i.e., along $Q_x$. Here, $\mathrm{rlu} = \lambda/2d$
is the reciprocal lattice unit, where $d$ is the corresponding lattice
plane distance. From the shape and the orientation of the GaN peak shown
in Fig.~\ref{fig:RSM}, we obtain a mosaic spread of $(1.9\pm0.2)^\circ$
(i.e., close to the FWHM of the $\omega$ scans) and a lateral correlation
length of 76~nm (close to the actual lateral size of the NWs, see
Fig.~\ref{fig:sem1}).

\begin{figure}[!t]
\includegraphics[width=6cm]{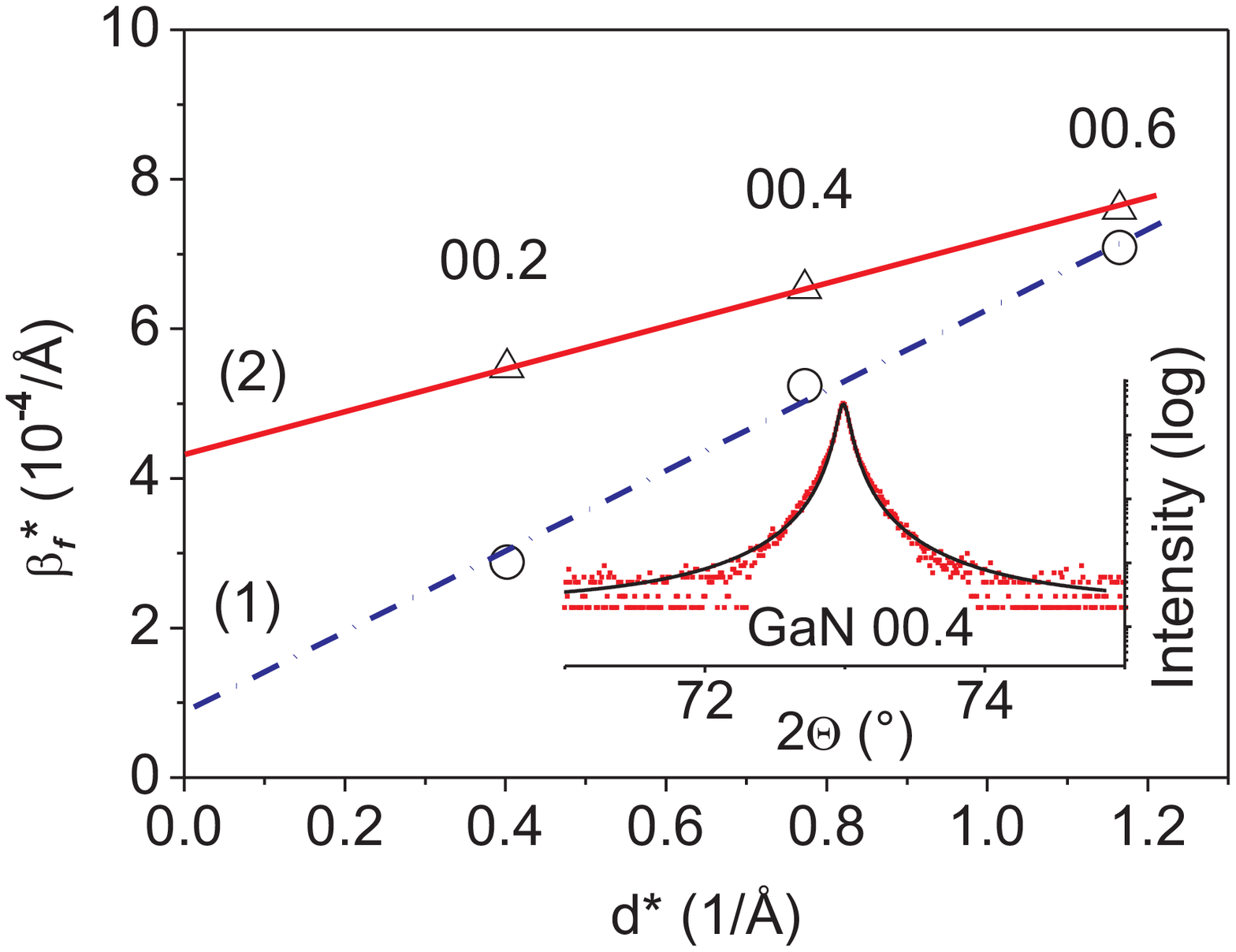}
\caption{(Color online) Williamson-Hall plot for the symmetrical GaN
reflections of samples~1~(2), see dashed line (solid line). The inset
shows the GaN 00.4 peak ($\omega/2\Theta$ scan) of sample 2  fitted with
a Lorentzian curve.}
\label{fig:WHplot}
\end{figure}

The widths of $\omega/2\Theta$ scans across the GaN 00.n reflections
with n$=2, 4, 6$ increase linearly with the reflection order n, which is
the result expected for broadening governed by micro-strain. The line shapes
are nearly Lorentzian, so that we obtain the breadth $\beta_{f}$ of the
physically broadened profile $\beta_{f}=\beta_{h}-\beta_{g}$ where
$\beta_{g}$ is the breadth of the apparatus function and $\beta_{h}$
that of the measured curve \cite{Klug1974}. Small asymmetries of the peaks
are neglected here. We assume that each GaN NW is a coherently diffracting
domain separated by air or a small-angle grain boundary in case of
coalescence. These domains are tilted with respect to the interface and
twisted around the surface normal by different angles reflected by broad
peaks in the $\omega$ and $\varphi$ scans. The $\omega/2\Theta$ scans,
instead, are broadened due to the finite size of the NWs and due to the
inhomogeneous deformations inside the NWs.
The integral breadth of the physically broadened line profile is
$\beta_{f}(2\Theta)= A/I_{0}$, where $A$ is the integral intensity and
$I_{0}$ the peak intensity of the corresponding maximum. We assume that
this breadth $\beta_{f}$ of our sample consists of size and strain
components $\beta_{S}$ and $\beta_{D}$. The size component $\beta_{S}$
does not depend on the diffraction angle. The strain component $\beta_{D}$
grows linearly with the order of reflection \cite{Klug1974}. We apply the
reciprocal lattice notation: $\left|d^{*}\right|= 1/d= 2\sin(\Theta)/\lambda$
is the absolute value of the reciprocal lattice vector and
$\beta_{f}^{*}=\beta_{f} \cos(\Theta)/\lambda$. For a Lorentzian peakshape
the two contributions have to be added \cite{langford}:
\begin{equation}
\beta_{f}^{*} = \beta_{S}^{*}+ \beta_{D}^{*} = \beta_{S}^{*} +2e d^{*},
\label{eq4}
\end{equation}
where $e = \beta_{D}^{*}/2 d^{*} = \Delta d/d$ is the strain.\cite{langford}
We can interprete this strain as the variation of $d$ spacing within the NWs.
The plot of $\beta_{f}^{*}$ over $d^{*}$ is expected to give a straight
line.\cite{williamson53,langford} Such a plot shown in Fig.~\ref{fig:WHplot}
is a Williamson-Hall plot in reciprocal-lattice representation.
The intersection with the ordinate axis allows us to determine
$\beta_{S}^{*}$, corresponding to the size $l = 1/\beta_{S}^{*}$.
This is the size of an average domain in the direction of the diffraction
vector, i.e., we essentially probe a length along the NWs. The characteristic
size is $l=1.3~(0.3)$~$\mu$m for sample~1 (2). For sample 1 this length is
comparable to the NW length, but for sample~2 it is lower by a factor of
five pointing to a smaller domain size inside the wires. The micro-strain
$e=\pm2.8\times10^{-4}$ ($\pm1.4\times10^{-4}$) in the GaN NWs of sample~1 (2)
is obtained from the slopes of the Williamson-Hall plots. These values
describe the fluctuations of the lattice plane distances in the NWs, not
their average. We estimate an error of $\pm1\times10^{-4}$ in the determination
of the micro-strain. The inhomogeneous micro-strain is caused by residual
defects in the GaN NWs like dislocations or even small angle boundaries
arising after coalescence of the NWs. \cite{Consonni2009}

The important result obtained here is the fact that the micro-strain for
sample~1 is indeed considerably larger than that for sample~2, just as the
PL linewidths. A quantitative comparison of these values to the linewidth
obtained in PL is, however, not straightforward. Coalescence presumably
generates a rather complex strain field, including shear strain, while we
only measure the $z$-component of the complete strain tensor. To give an
impression of the range of changes of the bandgap $\Delta E$ by a general
strain with $\varepsilon_{zz} =\pm1\times10^{-4}$, we follow
Ref.~\onlinecite{goshPRB2002} and obtain $\Delta E_b=1.65$~meV and
$\Delta E_h=5$~meV for pure biaxial ($\Delta E_b$) and pure hydrostatic
($\Delta E_h$) strain, respectively. Considering that at least part of the
linewidth is caused by the random distribution of donors and the resulting
energy dispersion, it is clear that only a fraction of the strain measured
by XRD manifests itself in the linewidth observed in PL. We believe that
this finding reflects the presence of dislocations in the coalescence
junctions. Dislocations are known to act as nonradiative recombination
centers and presumably suppress the PL from the junctions which act as
the dominant source of strain measured by XRD.

\section{Summary}
We have performed structural characterization
of MBE grown arrays of GaN NWs on a Si 111 substrate. The average lattice
in our GaN NWs is fully relaxed. Typical residual strains in the wires can
reach values from $\pm(0.015)\%$ to $\pm(0.03)\%$. The local distribution of strain is
visible in TEM micrographs. The observed micro-strain contributes to
broadening of the lines measured by photoluminescence spectroscopy.

\section{Acknowledgement}
The authors thank U.~Jahn for critical reading of the manuscript,
A.K. Bluhm for the scanning electron micrographs, A. Pfeiffer and D.
Steffen for the preparation of the TEM samples and C. Ch\'{e}ze,
V. Kaganer, and L. Geelhaar for helpful discussions.

\section{References}

\begin{thebibliography}{22}%
\makeatletter
\providecommand \@ifxundefined [1]{%
 \@ifx{#1\undefined}
}%
\providecommand \@ifnum [1]{%
 \ifnum #1\expandafter \@firstoftwo
 \else \expandafter \@secondoftwo
 \fi
}%
\providecommand \@ifx [1]{%
 \ifx #1\expandafter \@firstoftwo
 \else \expandafter \@secondoftwo
 \fi
}%
\providecommand \natexlab [1]{#1}%
\providecommand \enquote  [1]{``#1''}%
\providecommand \bibnamefont  [1]{#1}%
\providecommand \bibfnamefont [1]{#1}%
\providecommand \citenamefont [1]{#1}%
\providecommand \href@noop [0]{\@secondoftwo}%
\providecommand \href [0]{\begingroup \@sanitize@url \@href}%
\providecommand \@href[1]{\@@startlink{#1}\@@href}%
\providecommand \@@href[1]{\endgroup#1\@@endlink}%
\providecommand \@sanitize@url [0]{\catcode `\\12\catcode `\$12\catcode
  `\&12\catcode `\#12\catcode `\^12\catcode `\_12\catcode `\%12\relax}%
\providecommand \@@startlink[1]{}%
\providecommand \@@endlink[0]{}%
\providecommand \url  [0]{\begingroup\@sanitize@url \@url }%
\providecommand \@url [1]{\endgroup\@href {#1}{\urlprefix }}%
\providecommand \urlprefix  [0]{URL }%
\providecommand \Eprint [0]{\href }%
\providecommand \doibase [0]{http://dx.doi.org/}%
\providecommand \selectlanguage [0]{\@gobble}%
\providecommand \bibinfo  [0]{\@secondoftwo}%
\providecommand \bibfield  [0]{\@secondoftwo}%
\providecommand \translation [1]{[#1]}%
\providecommand \BibitemOpen [0]{}%
\providecommand \bibitemStop [0]{}%
\providecommand \bibitemNoStop [0]{.\EOS\space}%
\providecommand \EOS [0]{\spacefactor3000\relax}%
\providecommand \BibitemShut  [1]{\csname bibitem#1\endcsname}%
\let\auto@bib@innerbib\@empty
\bibitem [{\citenamefont {Harui}\ \emph {et~al.}(2008)\citenamefont {Harui},
  \citenamefont {Tamiya}, \citenamefont {Akagi}, \citenamefont {Miyake},
  \citenamefont {Hiramatsu}, \citenamefont {Araki},\ and\ \citenamefont
  {Nanishi}}]{harui_jjap_08}%
  \BibitemOpen
  \bibfield  {author} {\bibinfo {author} {\bibfnamefont {S.}~\bibnamefont
  {Harui}}, \bibinfo {author} {\bibfnamefont {H.}~\bibnamefont {Tamiya}},
  \bibinfo {author} {\bibfnamefont {T.}~\bibnamefont {Akagi}}, \bibinfo
  {author} {\bibfnamefont {H.}~\bibnamefont {Miyake}}, \bibinfo {author}
  {\bibfnamefont {K.}~\bibnamefont {Hiramatsu}}, \bibinfo {author}
  {\bibfnamefont {T.}~\bibnamefont {Araki}}, \ and\ \bibinfo {author}
  {\bibfnamefont {Y.}~\bibnamefont {Nanishi}},\ }\href@noop {} {\bibfield
  {journal} {\bibinfo  {journal} {Jap. J. Appl. Phys.}\ }\textbf {\bibinfo
  {volume} {47}},\ \bibinfo {pages} {5330} (\bibinfo {year}
  {2008})}\BibitemShut {NoStop}%
\bibitem [{\citenamefont {Consonni}\ \emph {et~al.}(2010)\citenamefont
  {Consonni}, \citenamefont {Knelangen}, \citenamefont {Geelhaar},
  \citenamefont {Trampert},\ and\ \citenamefont {Riechert}}]{consonni_prb_10}%
  \BibitemOpen
  \bibfield  {author} {\bibinfo {author} {\bibfnamefont {V.}~\bibnamefont
  {Consonni}}, \bibinfo {author} {\bibfnamefont {M.}~\bibnamefont {Knelangen}},
  \bibinfo {author} {\bibfnamefont {L.}~\bibnamefont {Geelhaar}}, \bibinfo
  {author} {\bibfnamefont {A.}~\bibnamefont {Trampert}}, \ and\ \bibinfo
  {author} {\bibfnamefont {H.}~\bibnamefont {Riechert}},\ }\href@noop {}
  {\bibfield  {journal} {\bibinfo  {journal} {Phys. Rev. B}\ }\textbf {\bibinfo
  {volume} {81}},\ \bibinfo {pages} {085310} (\bibinfo {year}
  {2010})}\BibitemShut {NoStop}%
\bibitem [{\citenamefont {Ertekin}\ \emph {et~al.}(2005)\citenamefont
  {Ertekin}, \citenamefont {Greaney}, \citenamefont {Chrzan},\ and\
  \citenamefont {Sands}}]{ertekin_jap_05}%
  \BibitemOpen
  \bibfield  {author} {\bibinfo {author} {\bibfnamefont {E.}~\bibnamefont
  {Ertekin}}, \bibinfo {author} {\bibfnamefont {P.~A.}\ \bibnamefont
  {Greaney}}, \bibinfo {author} {\bibfnamefont {D.~C.}\ \bibnamefont {Chrzan}},
  \ and\ \bibinfo {author} {\bibfnamefont {T.~D.}\ \bibnamefont {Sands}},\
  }\href@noop {} {\bibfield  {journal} {\bibinfo  {journal} {J. Appl. Phys.}\
  }\textbf {\bibinfo {volume} {97}},\ \bibinfo {pages} {114325} (\bibinfo
  {year} {2005})}\BibitemShut {NoStop}%
\bibitem [{\citenamefont {Glas}(2006)}]{glas_prb_06}%
  \BibitemOpen
  \bibfield  {author} {\bibinfo {author} {\bibfnamefont {F.}~\bibnamefont
  {Glas}},\ }\href@noop {} {\bibfield  {journal} {\bibinfo  {journal} {Phys.
  Rev. B}\ }\textbf {\bibinfo {volume} {74}},\ \bibinfo {pages} {121302(R)}
  (\bibinfo {year} {2006})}\BibitemShut {NoStop}%
\bibitem [{\citenamefont {Calleja}\ \emph {et~al.}(2000)\citenamefont
  {Calleja}, \citenamefont {S{\'a}nchez-Garc{\'i}a}, \citenamefont
  {S{\'a}nchez}, \citenamefont {Calle}, \citenamefont {Naranjo}, \citenamefont
  {Mu{\~n}oz}, \citenamefont {Jahn},\ and\ \citenamefont
  {Ploog}}]{calleja_prb_00}%
  \BibitemOpen
  \bibfield  {author} {\bibinfo {author} {\bibfnamefont {E.}~\bibnamefont
  {Calleja}}, \bibinfo {author} {\bibfnamefont {M.~A.}\ \bibnamefont
  {S{\'a}nchez-Garc{\'i}a}}, \bibinfo {author} {\bibfnamefont {F.~J.}\
  \bibnamefont {S{\'a}nchez}}, \bibinfo {author} {\bibfnamefont
  {F.}~\bibnamefont {Calle}}, \bibinfo {author} {\bibfnamefont {F.~B.}\
  \bibnamefont {Naranjo}}, \bibinfo {author} {\bibfnamefont {E.}~\bibnamefont
  {Mu{\~n}oz}}, \bibinfo {author} {\bibfnamefont {U.}~\bibnamefont {Jahn}}, \
  and\ \bibinfo {author} {\bibfnamefont {K.~H.}\ \bibnamefont {Ploog}},\
  }\href@noop {} {\bibfield  {journal} {\bibinfo  {journal} {Phys. Rev. B}\
  }\textbf {\bibinfo {volume} {62}},\ \bibinfo {pages} {16826} (\bibinfo {year}
  {2000})}\BibitemShut {NoStop}%
\bibitem [{\citenamefont {Brandt}\ \emph {et~al.}(2010)\citenamefont {Brandt},
  \citenamefont {Pf{\"u}ller}, \citenamefont {Ch\'{e}ze}, \citenamefont
  {Geelhaar},\ and\ \citenamefont {Riechert}}]{brandt2010}%
  \BibitemOpen
  \bibfield  {author} {\bibinfo {author} {\bibfnamefont {O.}~\bibnamefont
  {Brandt}}, \bibinfo {author} {\bibfnamefont {C.}~\bibnamefont {Pf{\"u}ller}},
  \bibinfo {author} {\bibfnamefont {C.}~\bibnamefont {Ch\'{e}ze}}, \bibinfo
  {author} {\bibfnamefont {L.}~\bibnamefont {Geelhaar}}, \ and\ \bibinfo
  {author} {\bibfnamefont {H.}~\bibnamefont {Riechert}},\ }\href@noop {}
  {\bibfield  {journal} {\bibinfo  {journal} {Phys. Rev. B}\ }\textbf {\bibinfo
  {volume} {81}},\ \bibinfo {pages} {045302} (\bibinfo {year}
  {2010})}\BibitemShut {NoStop}%
\bibitem [{\citenamefont {van Nostrand}\ \emph {et~al.}(2006)\citenamefont {van
  Nostrand}, \citenamefont {Averett}, \citenamefont {Cortez}, \citenamefont
  {Boeckl}, \citenamefont {Stutz}, \citenamefont {Sanford}, \citenamefont
  {Davydov},\ and\ \citenamefont {Albrecht}}]{nostrand_jcg_06}%
  \BibitemOpen
  \bibfield  {author} {\bibinfo {author} {\bibfnamefont {J.~E.}\ \bibnamefont
  {van Nostrand}}, \bibinfo {author} {\bibfnamefont {K.~L.}\ \bibnamefont
  {Averett}}, \bibinfo {author} {\bibfnamefont {R.}~\bibnamefont {Cortez}},
  \bibinfo {author} {\bibfnamefont {J.}~\bibnamefont {Boeckl}}, \bibinfo
  {author} {\bibfnamefont {C.~E.}\ \bibnamefont {Stutz}}, \bibinfo {author}
  {\bibfnamefont {N.~A.}\ \bibnamefont {Sanford}}, \bibinfo {author}
  {\bibfnamefont {A.~V.}\ \bibnamefont {Davydov}}, \ and\ \bibinfo {author}
  {\bibfnamefont {J.~D.}\ \bibnamefont {Albrecht}},\ }\href@noop {} {\bibfield
  {journal} {\bibinfo  {journal} {J. Cryst. Growth}\ }\textbf {\bibinfo
  {volume} {287}},\ \bibinfo {pages} {500} (\bibinfo {year}
  {2006})}\BibitemShut {NoStop}%
\bibitem [{\citenamefont {Furtmayr}\ \emph {et~al.}(2008)\citenamefont
  {Furtmayr}, \citenamefont {Vielemeyer}, \citenamefont {Stutzmann},
  \citenamefont {Laufer}, \citenamefont {Meyer},\ and\ \citenamefont
  {Eickhoff}}]{furtmayr_jap_08}%
  \BibitemOpen
  \bibfield  {author} {\bibinfo {author} {\bibfnamefont {F.}~\bibnamefont
  {Furtmayr}}, \bibinfo {author} {\bibfnamefont {M.}~\bibnamefont
  {Vielemeyer}}, \bibinfo {author} {\bibfnamefont {M.}~\bibnamefont
  {Stutzmann}}, \bibinfo {author} {\bibfnamefont {A.}~\bibnamefont {Laufer}},
  \bibinfo {author} {\bibfnamefont {B.~K.}\ \bibnamefont {Meyer}}, \ and\
  \bibinfo {author} {\bibfnamefont {M.}~\bibnamefont {Eickhoff}},\ }\href@noop
  {} {\bibfield  {journal} {\bibinfo  {journal} {J. Appl. Phys.}\ }\textbf
  {\bibinfo {volume} {104}},\ \bibinfo {pages} {074309} (\bibinfo {year}
  {2008})}\BibitemShut {NoStop}%
\bibitem [{\citenamefont {Corfdir}\ \emph {et~al.}(2009)\citenamefont
  {Corfdir}, \citenamefont {Lefebvre}, \citenamefont {Ristic}, \citenamefont
  {Valvin}, \citenamefont {Calleja}, \citenamefont {Trampert}, \citenamefont
  {Gani\`{e}re},\ and\ \citenamefont {Deveaud-Pl\'{e}dran}}]{corfdir_jap_09}%
  \BibitemOpen
  \bibfield  {author} {\bibinfo {author} {\bibfnamefont {P.}~\bibnamefont
  {Corfdir}}, \bibinfo {author} {\bibfnamefont {P.}~\bibnamefont {Lefebvre}},
  \bibinfo {author} {\bibfnamefont {J.}~\bibnamefont {Ristic}}, \bibinfo
  {author} {\bibfnamefont {P.}~\bibnamefont {Valvin}}, \bibinfo {author}
  {\bibfnamefont {E.}~\bibnamefont {Calleja}}, \bibinfo {author} {\bibfnamefont
  {A.}~\bibnamefont {Trampert}}, \bibinfo {author} {\bibfnamefont {J.-D.}\
  \bibnamefont {Gani\`{e}re}}, \ and\ \bibinfo {author} {\bibfnamefont
  {B.}~\bibnamefont {Deveaud-Pl\'{e}dran}},\ }\href@noop {} {\bibfield
  {journal} {\bibinfo  {journal} {J. Appl. Phys.}\ }\textbf {\bibinfo {volume}
  {105}},\ \bibinfo {pages} {013113} (\bibinfo {year} {2009})}\BibitemShut
  {NoStop}%
\bibitem [{\citenamefont {Dogan}\ \emph {et~al.}()\citenamefont {Dogan},
  \citenamefont {Brandt}, \citenamefont {Pf\"uller}, \citenamefont {Blum},
  \citenamefont {Geelhaar},\ and\ \citenamefont {Riechert}}]{Dogan2011}%
  \BibitemOpen
  \bibfield  {author} {\bibinfo {author} {\bibfnamefont {P.}~\bibnamefont
  {Dogan}}, \bibinfo {author} {\bibfnamefont {O.}~\bibnamefont {Brandt}},
  \bibinfo {author} {\bibfnamefont {C.}~\bibnamefont {Pf\"uller}}, \bibinfo
  {author} {\bibfnamefont {A.-K.}\ \bibnamefont {Blum}}, \bibinfo {author}
  {\bibfnamefont {L.}~\bibnamefont {Geelhaar}}, \ and\ \bibinfo {author}
  {\bibfnamefont {H.}~\bibnamefont {Riechert}},\ }\href@noop {} {\bibinfo
  {journal} {J. Cryst. Growth, doi:10.1016/j.jcrysgro.2010.12.081}\
  }\BibitemShut {NoStop}%
\bibitem [{\citenamefont {Fewster}\ and\ \citenamefont
  {Andrew}(1995)}]{Fewster95}%
  \BibitemOpen
\bibfield  {journal} {  }\bibfield  {author} {\bibinfo {author} {\bibfnamefont
  {P.}~\bibnamefont {Fewster}}\ and\ \bibinfo {author} {\bibfnamefont
  {N.}~\bibnamefont {Andrew}},\ }\href@noop {} {\bibfield  {journal} {\bibinfo
  {journal} {J. Appl. Cryst.}\ }\textbf {\bibinfo {volume} {28}},\ \bibinfo
  {pages} {451} (\bibinfo {year} {1995})}\BibitemShut {NoStop}%
\bibitem [{\citenamefont {Kornitzer}\ \emph {et~al.}(1999)\citenamefont
  {Kornitzer}, \citenamefont {Ebner}, \citenamefont {Grehl}, \citenamefont
  {Thonke}, \citenamefont {Sauer}, \citenamefont {Kirchner}, \citenamefont
  {Schwegler}, \citenamefont {Kamp}, \citenamefont {Leszczynski}, \citenamefont
  {Grzegory},\ and\ \citenamefont {Porowski}}]{kornitzer_pssb_99}%
  \BibitemOpen
  \bibfield  {author} {\bibinfo {author} {\bibfnamefont {K.}~\bibnamefont
  {Kornitzer}}, \bibinfo {author} {\bibfnamefont {T.}~\bibnamefont {Ebner}},
  \bibinfo {author} {\bibfnamefont {M.}~\bibnamefont {Grehl}}, \bibinfo
  {author} {\bibfnamefont {K.}~\bibnamefont {Thonke}}, \bibinfo {author}
  {\bibfnamefont {R.}~\bibnamefont {Sauer}}, \bibinfo {author} {\bibfnamefont
  {C.}~\bibnamefont {Kirchner}}, \bibinfo {author} {\bibfnamefont
  {V.}~\bibnamefont {Schwegler}}, \bibinfo {author} {\bibfnamefont
  {M.}~\bibnamefont {Kamp}}, \bibinfo {author} {\bibfnamefont {M.}~\bibnamefont
  {Leszczynski}}, \bibinfo {author} {\bibfnamefont {I.}~\bibnamefont
  {Grzegory}}, \ and\ \bibinfo {author} {\bibfnamefont {S.}~\bibnamefont
  {Porowski}},\ }\href@noop {} {\bibfield  {journal} {\bibinfo  {journal}
  {Physica Status Solidi B}\ }\textbf {\bibinfo {volume} {216}},\ \bibinfo
  {pages} {5} (\bibinfo {year} {1999})}\BibitemShut {NoStop}%
\bibitem [{\citenamefont {Wysmolek}\ \emph {et~al.}(2002)\citenamefont
  {Wysmolek}, \citenamefont {Korona}, \citenamefont {St{\c{e}}pniewski},
  \citenamefont {Baranowski}, \citenamefont {B{\l{}}oniarz}, \citenamefont
  {Potemski}, \citenamefont {Jones}, \citenamefont {Look}, \citenamefont
  {Kuhl}, \citenamefont {Park},\ and\ \citenamefont {Lee}}]{wysmolek_prb_02}%
  \BibitemOpen
  \bibfield  {author} {\bibinfo {author} {\bibfnamefont {A.}~\bibnamefont
  {Wysmolek}}, \bibinfo {author} {\bibfnamefont {K.~P.}\ \bibnamefont
  {Korona}}, \bibinfo {author} {\bibfnamefont {R.}~\bibnamefont
  {St{\c{e}}pniewski}}, \bibinfo {author} {\bibfnamefont {J.~M.}\ \bibnamefont
  {Baranowski}}, \bibinfo {author} {\bibfnamefont {J.}~\bibnamefont
  {B{\l{}}oniarz}}, \bibinfo {author} {\bibfnamefont {M.}~\bibnamefont
  {Potemski}}, \bibinfo {author} {\bibfnamefont {R.~L.}\ \bibnamefont {Jones}},
  \bibinfo {author} {\bibfnamefont {D.~C.}\ \bibnamefont {Look}}, \bibinfo
  {author} {\bibfnamefont {J.}~\bibnamefont {Kuhl}}, \bibinfo {author}
  {\bibfnamefont {S.~S.}\ \bibnamefont {Park}}, \ and\ \bibinfo {author}
  {\bibfnamefont {S.~K.}\ \bibnamefont {Lee}},\ }\href@noop {} {\bibfield
  {journal} {\bibinfo  {journal} {Phys. Rev. B}\ }\textbf {\bibinfo {volume}
  {66}},\ \bibinfo {pages} {245317} (\bibinfo {year} {2002})}\BibitemShut
  {NoStop}%
\bibitem [{\citenamefont {Consonni}\ \emph {et~al.}(2009)\citenamefont
  {Consonni}, \citenamefont {Knelangen}, \citenamefont {Jahn}, \citenamefont
  {Trampert}, \citenamefont {Geelhaar},\ and\ \citenamefont
  {Riechert}}]{Consonni2009}%
  \BibitemOpen
  \bibfield  {author} {\bibinfo {author} {\bibfnamefont {V.}~\bibnamefont
  {Consonni}}, \bibinfo {author} {\bibfnamefont {M.}~\bibnamefont {Knelangen}},
  \bibinfo {author} {\bibfnamefont {U.}~\bibnamefont {Jahn}}, \bibinfo {author}
  {\bibfnamefont {A.}~\bibnamefont {Trampert}}, \bibinfo {author}
  {\bibfnamefont {L.}~\bibnamefont {Geelhaar}}, \ and\ \bibinfo {author}
  {\bibfnamefont {H.}~\bibnamefont {Riechert}},\ }\href@noop {} {\bibfield
  {journal} {\bibinfo  {journal} {Appl. Phys. Lett.}\ }\textbf {\bibinfo
  {volume} {95}},\ \bibinfo {pages} {241910} (\bibinfo {year}
  {2009})}\BibitemShut {NoStop}%
\bibitem [{\citenamefont {Sun}\ \emph {et~al.}(2002)\citenamefont {Sun},
  \citenamefont {Brandt}, \citenamefont {Liu}, \citenamefont {Trampert},
  \citenamefont {Ploog}, \citenamefont {Bl\"asing},\ and\ \citenamefont
  {Krost}}]{Sun2002}%
  \BibitemOpen
  \bibfield  {author} {\bibinfo {author} {\bibfnamefont {Y.~J.}\ \bibnamefont
  {Sun}}, \bibinfo {author} {\bibfnamefont {O.}~\bibnamefont {Brandt}},
  \bibinfo {author} {\bibfnamefont {T.~Y.}\ \bibnamefont {Liu}}, \bibinfo
  {author} {\bibfnamefont {A.}~\bibnamefont {Trampert}}, \bibinfo {author}
  {\bibfnamefont {K.~H.}\ \bibnamefont {Ploog}}, \bibinfo {author}
  {\bibfnamefont {J.}~\bibnamefont {Bl\"asing}}, \ and\ \bibinfo {author}
  {\bibfnamefont {A.}~\bibnamefont {Krost}},\ }\href@noop {} {\bibfield
  {journal} {\bibinfo  {journal} {Appl. Phys. Lett.}\ }\textbf {\bibinfo
  {volume} {81}},\ \bibinfo {pages} {4928} (\bibinfo {year}
  {2002})}\BibitemShut {NoStop}%
\bibitem [{\citenamefont {Moram}\ and\ \citenamefont
  {Vickers}(2009)}]{moram_rpp_09}%
  \BibitemOpen
  \bibfield  {author} {\bibinfo {author} {\bibfnamefont {M.~A.}\ \bibnamefont
  {Moram}}\ and\ \bibinfo {author} {\bibfnamefont {M.~E.}\ \bibnamefont
  {Vickers}},\ }\href@noop {} {\bibfield  {journal} {\bibinfo  {journal} {Rep.
  Prog. Phys.}\ }\textbf {\bibinfo {volume} {72}},\ \bibinfo {pages} {036502}
  (\bibinfo {year} {2009})}\BibitemShut {NoStop}%
\bibitem [{\citenamefont {Stepanov}(1997)}]{Stepanov1997}%
  \BibitemOpen
  \bibfield  {author} {\bibinfo {author} {\bibfnamefont {S.}~\bibnamefont
  {Stepanov}},\ }\href@noop {} {\emph {\bibinfo {title} {Collection of
  software}}}\ (\bibinfo  {publisher} {Sergey Stepanov's X-ray server,
  http://sergey.gmca.aps.anl.gov/},\ \bibinfo {address} {Chicago},\ \bibinfo
  {year} {1997})\BibitemShut {NoStop}%
\bibitem [{\citenamefont {Robins}\ \emph {et~al.}(2007)\citenamefont {Robins},
  \citenamefont {Bertness}, \citenamefont {Barker}, \citenamefont {Sanford},\
  and\ \citenamefont {Schlager}}]{Robins2007}%
  \BibitemOpen
  \bibfield  {author} {\bibinfo {author} {\bibfnamefont {L.~H.}\ \bibnamefont
  {Robins}}, \bibinfo {author} {\bibfnamefont {K.~A.}\ \bibnamefont
  {Bertness}}, \bibinfo {author} {\bibfnamefont {J.~M.}\ \bibnamefont
  {Barker}}, \bibinfo {author} {\bibfnamefont {N.~A.}\ \bibnamefont {Sanford}},
  \ and\ \bibinfo {author} {\bibfnamefont {J.~B.}\ \bibnamefont {Schlager}},\
  }\href@noop {} {\bibfield  {journal} {\bibinfo  {journal} {J. Appl. Phys.}\
  }\textbf {\bibinfo {volume} {101}},\ \bibinfo {pages} {113505} (\bibinfo
  {year} {2007})}\BibitemShut {NoStop}%
\bibitem [{\citenamefont {Klug}\ and\ \citenamefont
  {Alexander}(1974)}]{Klug1974}%
  \BibitemOpen
  \bibfield  {author} {\bibinfo {author} {\bibfnamefont {H.~P.}\ \bibnamefont
  {Klug}}\ and\ \bibinfo {author} {\bibfnamefont {L.~E.}\ \bibnamefont
  {Alexander}},\ }\href@noop {} {\emph {\bibinfo {title} {X-Ray Diffraction
  procedures}}}\ (\bibinfo  {publisher} {John Wiley},\ \bibinfo {address} {New
  York},\ \bibinfo {year} {1974})\BibitemShut {NoStop}%
\bibitem [{\citenamefont {Langford}(2000)}]{langford}%
  \BibitemOpen
  \bibfield  {author} {\bibinfo {author} {\bibfnamefont {J.~I.}\ \bibnamefont
  {Langford}},\ }\enquote {\bibinfo {title} {Industrial applications of x-ray
  diffraction},}\ \ (\bibinfo  {publisher} {Marcel Dekker Inc.},\ \bibinfo
  {address} {New York},\ \bibinfo {year} {2000})\ p.\ \bibinfo {pages}
  {751}\BibitemShut {NoStop}%
\bibitem [{\citenamefont {Williamson}\ and\ \citenamefont
  {Hall}(1953)}]{williamson53}%
  \BibitemOpen
  \bibfield  {author} {\bibinfo {author} {\bibfnamefont {G.~K.}\ \bibnamefont
  {Williamson}}\ and\ \bibinfo {author} {\bibfnamefont {W.~H.}\ \bibnamefont
  {Hall}},\ }\href@noop {} {\bibfield  {journal} {\bibinfo  {journal} {Acta
  Met.}\ }\textbf {\bibinfo {volume} {1}},\ \bibinfo {pages} {22} (\bibinfo
  {year} {1953})}\BibitemShut {NoStop}%
\bibitem [{\citenamefont {Gosh}\ \emph {et~al.}(2002)\citenamefont {Gosh},
  \citenamefont {Waltereit}, \citenamefont {Brandt}, \citenamefont {Grahn},\
  and\ \citenamefont {Ploog}}]{goshPRB2002}%
  \BibitemOpen
  \bibfield  {author} {\bibinfo {author} {\bibfnamefont {S.}~\bibnamefont
  {Gosh}}, \bibinfo {author} {\bibfnamefont {P.}~\bibnamefont {Waltereit}},
  \bibinfo {author} {\bibfnamefont {O.}~\bibnamefont {Brandt}}, \bibinfo
  {author} {\bibfnamefont {H.~T.}\ \bibnamefont {Grahn}}, \ and\ \bibinfo
  {author} {\bibfnamefont {K.~H.}\ \bibnamefont {Ploog}},\ }\href@noop {}
  {\bibfield  {journal} {\bibinfo  {journal} {Phys. Rev. B}\ }\textbf {\bibinfo
  {volume} {65}},\ \bibinfo {pages} {075202} (\bibinfo {year}
  {2002})}\BibitemShut {NoStop}%
\end{thebibliography}
%

\end{document}